\documentclass[RNAAS]{aastex631}

\usepackage{xurl}
\begin{document}

\title{Deep Learning for Astrophysics: An Open Textbook from the NASA Cosmic Origins AI/ML Science and Technology Interest Group}

\author{Yuan-Sen Ting}
\altaffiliation{Co-chair, NASA Cosmic Origins AI/ML STIG}
\affiliation{Department of Astronomy, The Ohio State University, Columbus, OH 43210, USA}
\affiliation{Center for Cosmology and AstroParticle Physics (CCAPP), The Ohio State University, Columbus, OH 43210, USA}
\affiliation{Max-Planck-Institut f\"ur Astronomie, K\"onigstuhl 17, D-69117 Heidelberg, Germany}

\author{Digvijay Wadekar}
\altaffiliation{Co-chair, NASA Cosmic Origins AI/ML STIG}
\affiliation{Center for Gravitational Physics, Department of Physics, The University of Texas at Austin, Austin, TX 78712, USA}

\author{Phillip Cargile}
\affiliation{Center for Astrophysics \textbar{} Harvard \& Smithsonian, 60 Garden Street, Cambridge, MA 02138, USA}

\author{Carol Cuesta-Lazaro}
\altaffiliation{Leadership Council, NASA Cosmic Origins AI/ML STIG}
\affiliation{Institute for Advanced Study, Princeton, NJ 08540, USA}
\affiliation{Center for Computational Astrophysics, Flatiron Institute, New York, NY 10010, USA}

\author{Andr\'e Curtis-Trudel}
\affiliation{Department of Philosophy, University of Cincinnati, Cincinnati, OH 45221, USA}

\author{Gregory Green}
\affiliation{School of Science, Westlake University, Hangzhou, Zhejiang 310030, China}

\author{Ryan McClelland}
\affiliation{NASA Goddard Space Flight Center, Greenbelt, MD 20771, USA}

\author{Daniel Muthukrishna}
\affiliation{MIT Kavli Institute for Astrophysics and Space Research, Massachusetts Institute of Technology, Cambridge, MA 02139, USA}
\affiliation{Center for Astrophysics \textbar{} Harvard \& Smithsonian, 60 Garden Street, Cambridge, MA 02138, USA}

\author{Tri Nguyen}
\affiliation{Center for Interdisciplinary Exploration and Research in Astrophysics (CIERA), Northwestern University, Evanston, IL 60201, USA}
\affiliation{Department of Astronomy and Astrophysics, University of Toronto, 50 St.\ George Street, Toronto, ON M5S 3H4, Canada}

\author{Helen Qu}
\affiliation{Center for Computational Astrophysics, Flatiron Institute, 162 Fifth Avenue, New York, NY 10010, USA}

\author{Tomasz Rozanski}
\affiliation{Research School of Astronomy and Astrophysics, The Australian National University, Canberra, ACT 2611, Australia}

\author{Anna Scaife}
\affiliation{Jodrell Bank Centre for Astrophysics, Department of Physics and Astronomy, University of Manchester, Manchester M13 9PL, UK}

\author{Jesse Thaler}
\affiliation{Center for Theoretical Physics, Massachusetts Institute of Technology, Cambridge, MA 02139, USA}
\affiliation{The NSF AI Institute for Artificial Intelligence and Fundamental Interactions (IAIFI), Massachusetts Institute of Technology, Cambridge, MA 02139, USA}

\author{Licia Verde}
\affiliation{ICREA and Institut de Ci\`encies del Cosmos (ICCUB), Universitat de Barcelona, Mart\'i i Franqu\`es 1, 08028 Barcelona, Spain}

\author{Francisco Villaescusa-Navarro}
\affiliation{Center for Computational Astrophysics, Flatiron Institute, 162 Fifth Avenue, New York, NY 10010, USA}

\author{John F. Wu}
\affiliation{Space Telescope Science Institute, 3700 San Martin Drive, Baltimore, MD 21218, USA}

\author{Duo Xu}
\affiliation{Canadian Institute for Theoretical Astrophysics, University of Toronto, 60 St.\ George Street, Toronto, ON M5S 3H8, Canada}

\author{Siyu Yao}
\affiliation{Department of Philosophy, School of Humanities, Shanghai Jiao Tong University, Shanghai, China}

\author{Alex Gagliano}
\altaffiliation{Leadership Council, NASA Cosmic Origins AI/ML STIG}
\affiliation{The NSF AI Institute for Artificial Intelligence and Fundamental Interactions (IAIFI), Massachusetts Institute of Technology, Cambridge, MA 02139, USA}

\author{Siddharth Mishra-Sharma}
\altaffiliation{Leadership Council, NASA Cosmic Origins AI/ML STIG}
\affiliation{Faculty of Computing and Data Sciences and Department of Physics, Boston University, Boston, MA 02215, USA}

\author{Andrew~K.~Saydjari}
\altaffiliation{Hubble Fellow}
\altaffiliation{Leadership Council, NASA Cosmic Origins AI/ML STIG}
\affiliation{Department of Astrophysical Sciences, Princeton University, Princeton, NJ 08544, USA}

\author{Georgios Valogiannis}
\altaffiliation{Leadership Council, NASA Cosmic Origins AI/ML STIG}
\affiliation{Department of Astronomy and Astrophysics and Kavli Institute for Cosmological Physics, University of Chicago, Chicago, IL 60637, USA}

\author{Peter Kurczynski}
\altaffiliation{Chief Scientist, NASA Cosmic Origins Program Office}
\affiliation{NASA Goddard Space Flight Center, Greenbelt, MD 20771, USA}

\author{Swara Ravindranath}
\altaffiliation{Deputy Chief Scientist, NASA Cosmic Origins Program Office}
\affiliation{Center for Research and Exploration in Space Science and Technology II (CRESST II), The Catholic University of America, Washington, DC 20064, USA}
\affiliation{NASA Goddard Space Flight Center, Greenbelt, MD 20771, USA}

\begin{abstract}
Recent community assessments identify education as a principal barrier to adopting modern machine learning in astronomy. We present \textit{Deep Learning for Astrophysics}, a freely available textbook at \url{https://deeplearning4astro.com}, curated from the NASA Cosmic Origins Artificial Intelligence and Machine Learning Science and Technology Interest Group (AI/ML STIG) lecture series. The book collects 23 chapters by 17 lecturers across six parts, moving from computational foundations and deep-learning architectures through generative modeling, simulation-based inference, reinforcement learning, and large-language-model agents to the practice of AI-laden science. Many include executable notebooks using astronomical data.
\end{abstract}

\section{The gap is adoption, not availability}

The scientific case for machine learning in astronomy is established. Data from the Vera C. Rubin Observatory and the Nancy Grace Roman Space Telescope will strain traditional analysis, and learned methods already aid classification and inference. Uptake remains the limiting problem. The U.S. NSF AI+MPS community paper \citep{Ferguson2026} lists AI education among strategic priorities for the mathematical and physical sciences. It names an undertrained community, distrust of opaque methods, and resources written by and for computer scientists as barriers, and proposes modular, stackable training in response.

A review of deep learning in astrophysics \citep{Ting2026} makes a complementary point. The value of these methods rests on understanding why they work, on architectural choices that encode physical assumptions and on testability, rather than on predictive accuracy alone. Understanding is the prerequisite for trust, and trust for adoption. This matters because AI can produce an appearance of understanding without the substance of it \citep{Messeri2024}.

Literacy matters even for those who never train a model. Astronomers referee papers, sit on panels, and edit journals, and as AI-generated work enters the literature, understanding the methods is what lets them separate scientific insight from plausible output. Both assessments point to the same response: domain scientists should build AI training tailored to their field, with domain-relevant examples and criteria for when a learned method is the right tool.

\section{Textbook design}

NASA's Cosmic Origins program supports the community studying how galaxies, stars, and cosmic structure form and evolve. The AI/ML STIG is a Science and Technology Interest Group of its Program Analysis Group (COPAG), open to the national and international community.\footnote{STIG website at \url{https://ai4astro.org/}.} The textbook grew from the premise that training the community can increase the science return of NASA's missions, and that the training must be domain-specific and time-efficient.

The book responds to obstacles named in the community paper and the group's founding proposal: resistance to AI among established researchers, the difficulty of assessing interdisciplinary skill, faculty time constraints that make full courses impractical, fragmented resources that lack astronomical context, and a pace of change that outruns standard curricula. The response is a training sequence assembled from stackable short modules, each covering a self-contained competency. The aim is not to turn astronomers into machine-learning engineers but to help researchers connect astronomical questions with machine-learning methods.\footnote{The author list is ordered as co-chairs, lecturers, leadership council, and Cosmic Origins program officers, with each group alphabetical by surname. Yuan-Sen Ting is both a co-chair and a lecturer, and Carol Cuesta-Lazaro is both a lecturer and a member of the leadership council.}

\section{The textbook}

\begin{figure}[t]
\centering
\includegraphics[width=\linewidth]{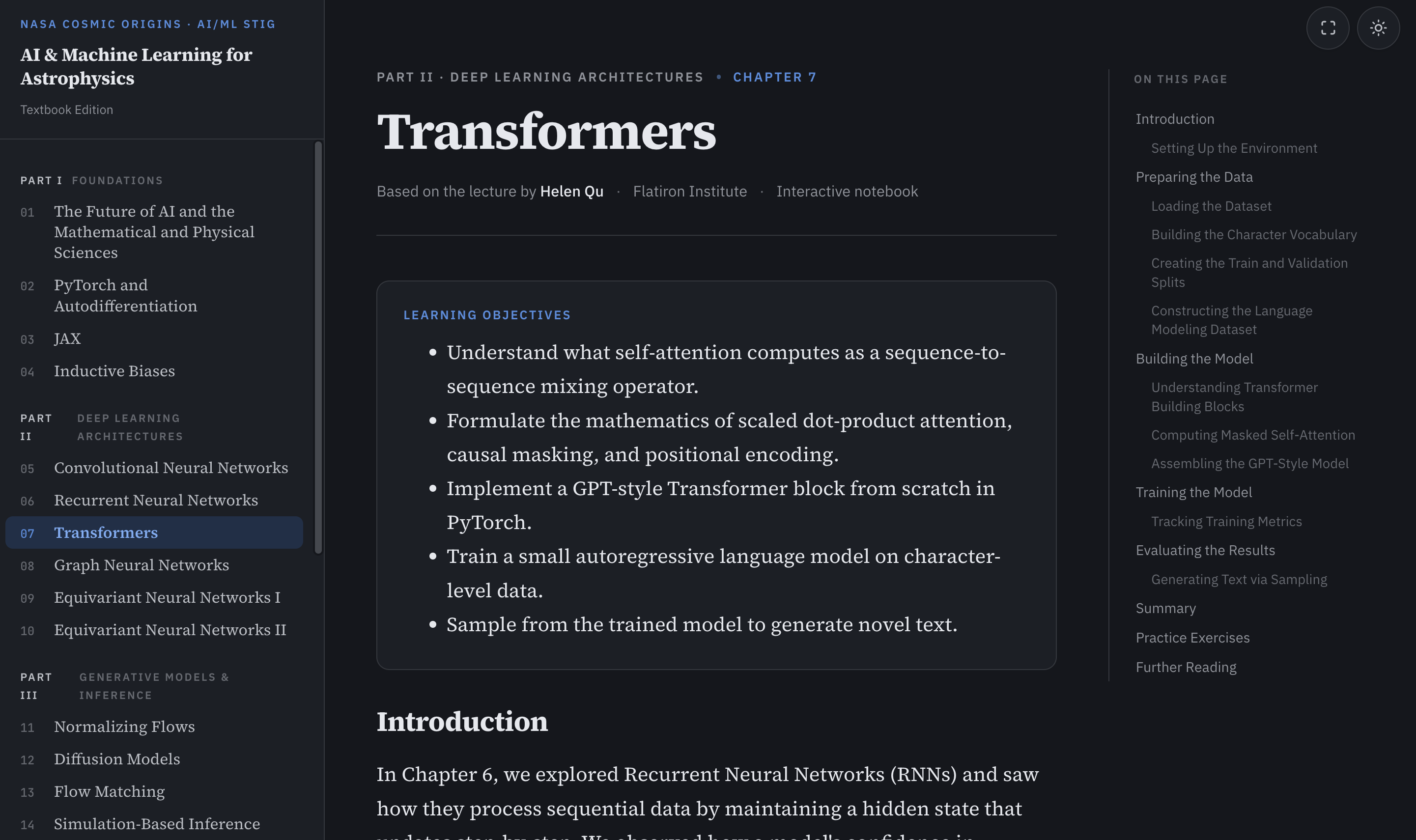}
\caption{The web version of \textit{Deep Learning for Astrophysics}. The textbook turns the AI/ML STIG lecture series into modular, domain-specific chapters for astronomers learning modern machine-learning methods.\label{fig:reader}}
\end{figure}

Over the past academic year the group ran an online lecture series, with recordings and materials collected on the STIG website.\footnote{\url{https://ai4astro.org/}.} We have curated these lectures into a single textbook, \textit{Deep Learning for Astrophysics}, available online.\footnote{\url{https://deeplearning4astro.com}.} The lectures, notebooks, and slides were edited into a consistent chapter format and reviewed by the contributing lecturers. The executable notebooks and their outputs are preserved, so readers can run and modify the methods rather than only read about them (Figure~\ref{fig:reader}). The exercises use real astronomical data, such as galaxy images, radio-galaxy morphologies, transient light curves, and stellar streams, rather than toy examples.

Each notebook chapter pairs the minimum theory a method needs with a worked astronomical example, so that its uses and its failure modes are learned together. Of the 23 chapters, seventeen are executable notebooks and six are concise readings, grouped into six parts. The intended reader is a graduate student or researcher who wants working knowledge rather than a survey. Because the parts are self-contained, the book can be read end to end or used as standalone modules in a course. The broader lecture series also included sessions on the AI funding landscape and on AI-enabled mission concepts; the textbook focuses on the technical and interpretive material most suited to chapter format.

\vspace{1.5\baselineskip}
\medskip
{\small
\noindent\textbf{Part I. Foundations}
\begin{itemize}\setlength\itemsep{0pt}
\item The Future of AI and the Mathematical and Physical Sciences \textit{(Jesse Thaler)}
\item PyTorch and Autodifferentiation \textit{(Phillip Cargile)}
\item JAX \textit{(Phillip Cargile)}
\item Inductive Biases \textit{(John F. Wu)}
\end{itemize}
\noindent\textbf{Part II. Deep Learning Architectures}
\begin{itemize}\setlength\itemsep{0pt}
\item Convolutional Neural Networks \textit{(John F. Wu)}
\item Recurrent Neural Networks \textit{(Daniel Muthukrishna)}
\item Transformers \textit{(Helen Qu)}
\item Graph Neural Networks \textit{(Tri Nguyen)}
\item Equivariant Neural Networks \textit{(Anna Scaife)}
\end{itemize}
\noindent\textbf{Part III. Generative Models \& Inference}
\begin{itemize}\setlength\itemsep{0pt}
\item Normalizing Flows \textit{(Gregory Green)}
\item Diffusion Models \textit{(Duo Xu)}
\item Flow Matching \textit{(Tomasz Rozanski)}
\item Simulation-Based Inference \textit{(Tomasz Rozanski)}
\end{itemize}
\noindent\textbf{Part IV. Reinforcement Learning}
\begin{itemize}\setlength\itemsep{0pt}
\item Reinforcement Learning \textit{(Carol Cuesta-Lazaro)}
\end{itemize}
\noindent\textbf{Part V. Large Language Models \& Agents}
\begin{itemize}\setlength\itemsep{0pt}
\item LLM API Basics \textit{(Yuan-Sen Ting)}
\item RAG and Function Tools \textit{(Yuan-Sen Ting)}
\item Model Context Protocol \textit{(Yuan-Sen Ting)}
\item LLM as Agent \textit{(Francisco Villaescusa-Navarro)}
\end{itemize}
\noindent\textbf{Part VI. AI, Science \& Society}
\begin{itemize}\setlength\itemsep{0pt}
\item From Text to Spaceship \textit{(Ryan McClelland)}
\item AI and Scientific Publishing \textit{(Licia Verde)}
\item Understanding in AI-Laden Science \textit{(Siyu Yao \& Andr\'e Curtis-Trudel)}
\end{itemize}
}

The textbook will continue to incorporate corrections and contributions from the community. It records the lecture series as a coherent path through AI in astronomy and a practical entry point for researchers.

\begin{acknowledgments}
The textbook was produced by editing the original lecture materials into a consistent chapter format with substantial assistance from large language models (Anthropic's Claude Opus 4.8 and OpenAI's GPT-5.5), which were also used to copy-edit this note. In both cases the text was checked and verified by the authors. We thank the Cosmic Origins Program Office and the COPAG Executive Committee for their support, and the community members whose questions shaped the lectures.
\end{acknowledgments}

\bibliography{Manuscript}
\bibliographystyle{aasjournal}

\end{document}